\documentclass[reprint,
 amsmath,amssymb,
 aps,
]{revtex4-2}
\usepackage{siunitx}
\usepackage{graphicx}
\usepackage{dcolumn}
\usepackage{bm}

\begin{document}

\preprint{APS/123-QED}

\title{Magic Velocity Selection in Atom Interferometry}
\author{Yuno Iwasaki}
\email{yiwasaki@berkeley.edu}

\author{Jack Roth}
\author{Madeline Bernstein}
\author{Andrew Christensen}
\author{Holger Mueller}
\altaffiliation[Also at ]{%
 Lawrence Berkeley National Laboratory, Berkeley, California 94720, USA
}
\affiliation{%
Department of Physics, University of California, Berkeley \\ Berkeley, California 94720, USA
}

\date{\today}

\begin{abstract}
Velocity-selective Raman transitions are widely used in atom interferometers to prepare atomic
ensembles with narrowly defined momentum distributions. However, differential light shifts between atomic energy levels generate velocity distributions that are correlated with the Raman beam intensity, and therefore with the position of the atoms within the laser beam. We show that these spatially inhomogeneous  velocity distributions interact with detuning-dependent systematic effects in a Bragg diffraction-based simultaneous conjugate Ramsey-Bord\'e interferometer. These interactions can induce systematic phase shifts of order 10 milliradians in the interferometer phase. We further identify a ``magic'' detuning for velocity selection and show that operating at this detuning suppresses the systematic phase shift. Magic velocity selection eliminates systematic errors arising from correlations between atom velocity and position, facilitating high-resolution atom interferometry experiments targeting sub-part-per-billion accuracy.
\end{abstract}


\maketitle


\section{\label{sec:level1}Introduction}
\begin{figure*}
    \centering
\includegraphics[width=\textwidth]{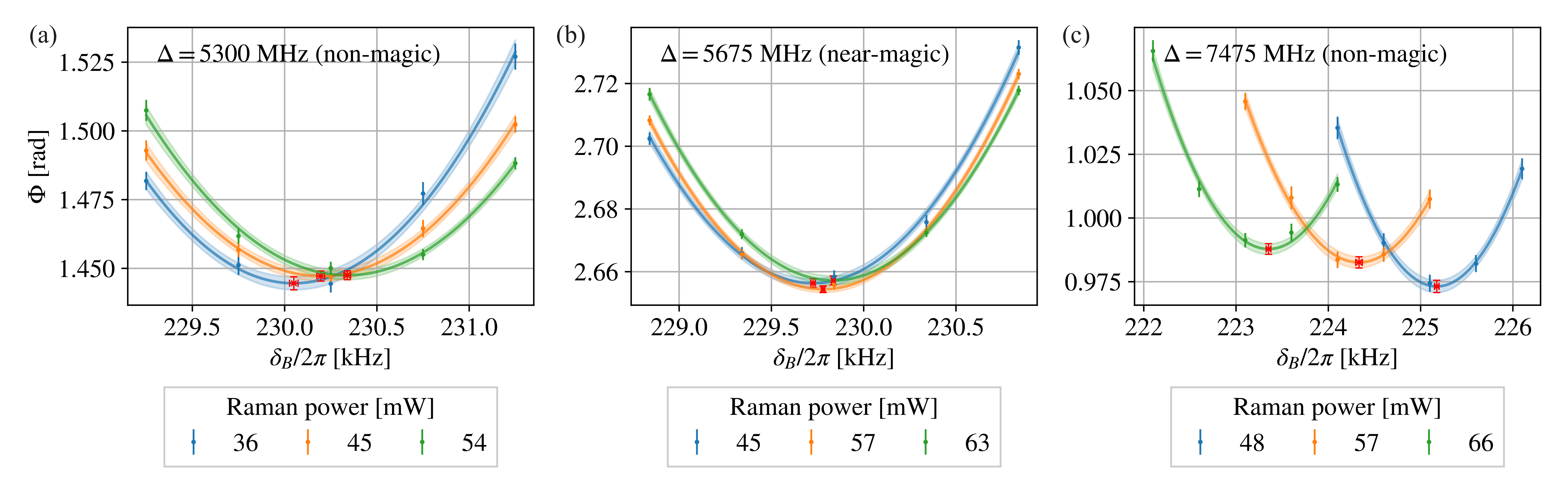}
    \caption{Interferometer phase $\Phi$ as a function of $\delta_B$, the relative frequency difference between the two beams used for Bragg diffraction. The data are shown for several combinations of Raman single-photon detuning $\Delta$ and Raman laser power. Red markers indicate the minima extracted from parabolic fits to the data. Panels (a), (b), and (c) show data taken at $\Delta = 5300~\mathrm{MHz}$, $5675~\mathrm{MHz},$ and $7475~\mathrm{MHz}$, respectively. Different colors denote different Raman powers used during velocity selection. From (b), we infer that $\mathrm{\Delta = 5675~MHz}$ is close to the magic detuning, as the three parabolas nearly coincide. In contrast, (a) and (c) indicate that $\mathrm{5300~MHz}$ and $\mathrm{7475~MHz}$ are blue- and red-detuned, respectively, relative to the magic detuning, because the three parabolas are horizontally shifted with respect to one another. Furthermore,  (c) shows that the minimum interferometer phase depends on the Raman power and is not recovered by scanning $\delta_B$.}
    \label{fig:testplot}
\end{figure*}
In atom interferometers, laser pulses are used to split atomic matter-waves into quantum superpositions that travel along different trajectories. When the paths are recombined, the accumulated phase difference between them can be measured by reading out the interference signal. Atom interferometry has been used in wide-reaching applications in fundamental physics, including measuring the gravitational constant \cite{Rosi_2014, kasevich_g_measurement}, testing the equivalence principle \cite{asenbaum_2020, He2023}, constraining dark matter and dark energy theories \cite{berkeley_alpha, Morel2020, Hamilton_2015}, and measuring the fine structure constant \cite{ berkeley_alpha, Morel2020}.

 Bragg diffraction enables multi-photon beam splitters, which increases the momentum separation between the interferometer arms. This results in a larger interferometric phase, enhancing the sensitivity of the interferometer \cite{Kovachy2015, beguin_2023}. However, it introduces the diffraction phase, an additional systematic effect \cite{estey_2015, gupta_three_path, Jamison_2014}. The diffraction phase is sensitive to the atom velocity \cite{estey_2015}. Therefore, accurately characterizing the diffraction phase requires understanding how the initial atomic velocity distribution is prepared. Velocity-selective Raman transitions are commonly used during state preparation to narrow the atomic velocity distribution \cite{vs_moler, kasevich_1991}. Unfortunately, the selected velocity depends on the differential light shift \cite{ge_2023, xu_2026}. Because the Raman laser intensity (and thus the light shift) varies across the beam, the selected velocity becomes spatially dependent, leading to a spatially inhomogeneous final velocity distribution \cite{PhysRevA.78.043615}. This makes the diffraction phase difficult to characterize, because the measured phase is a detection-efficiency-weighted average over atoms that acquire different velocity-dependent diffraction phases. The situation is further complicated by the dependence of the diffraction phase on the local Bragg beam intensity, which also varies across the atomic ensemble \cite{estey_2015, gupta_three_path, Jamison_2014}. Consequently, estimating the diffraction phase requires knowledge of the spatial velocity distribution of the atomic ensemble, rather than only its overall velocity spread. This complicates accurate correction for the diffraction phase.

In this work, we experimentally investigate the impact of the laser intensity during velocity selection on the diffraction phase in a simultaneous conjugate Ramsey-Bord\'e cesium atom interferometer using 4th-order Bragg beam splitters. By varying the single-photon detuning (i.e., the detuning of the Raman lasers from the intermediate excited state) and the intensity of the velocity-selective Raman pulses, we tune the spatial velocity variation imprinted on the atomic ensemble and measure the resulting  interferometer phase. As shown in Fig.\,\ref{fig:testplot}, we observe that the measured interferometer phase, $\Phi$, generally depends on the Raman laser intensity.  Furthermore, we identify a ``magic'' single-photon detuning for velocity selection at which $\Phi$ is insensitive to Raman laser intensity variations, as shown in Fig.\,\ref{fig:testplot}(b). Operating at this detuning eliminates correlations between the selected velocity and position, simplifying the systematic error due to diffraction phase. 

This paper is organized as follows. In Section \ref{subsec:apparatus}, we present an overview of the experimental apparatus. In Section \ref{subsec:vs_theory}, we review the dynamics of velocity-selective Raman transitions and show how a non-zero differential light shift induces local variations in the resonant velocity class. We then outline the conditions required to suppress this variation---namely, operating at a single-photon detuning where the differential light shift vanishes. Section \ref{subsec:bragg_diffraction} outlines the dynamics of Bragg diffraction. We compare the relative magnitudes of the expected signal phase and the anomalous phase contributions arising from the Bragg beam splitters. We present the results of full interferometer simulations to illustrate the dependence of the diffraction phase on laser intensity and atomic velocity. In Section \ref{sec:methods}, we describe the experimental procedure used to investigate this dependence. In Section \ref{sec:results}, we present our results. In Section \ref{sec:mwl_validation}, we corroborate the location of the velocity selection magic detuning using two independent experimental methods.
\section{\label{sec:background} Background}
\subsection{\label{subsec:apparatus}Experimental Apparatus}
The apparatus is a cesium atomic fountain interferometer in a $\mathrm{5\text{-}meter}$ tall vacuum chamber. The experimental sequence begins with a 2-D magneto-optical trap (MOT), which loads a 3-D MOT to perform a moving molasses launch at $\mathrm{6.2~m/s}$. As the atoms are released, polarization gradient cooling is applied, followed by Raman sideband cooling in the moving frame during the launch \cite{rsc_ref}. This reduces the atomic velocity distribution to a width of approximately $\numrange{2}{3}$ recoil velocities ($v_r\approx\mathrm{3.5~mm/s}$ for cesium) and leaves atoms mostly in the $F=3, m_F=3$ ground state.  An adiabatic rapid passage microwave pulse then transfers the atoms to the $m_F=0$ state, after which two consecutive velocity-selective Raman $\pi$ pulses are applied (``double velocity selection''). Resonant 
blow-away pulses after each Raman pulse remove the untransferred atomic populations. This prepares  the atoms in the $F=3, m_F = 0$ ground hyperfine state and further narrows the width of the velocity distribution to $\sim\!0.1\,v_r$. Finally, interferometry is performed using Bragg diffraction and Bloch oscillations (see Fig.\,4).

After the interferometer sequence, the output populations are measured by allowing the atoms to fall through an on-resonance light sheet. Because the different output momentum states reach the light sheet at different times, their fluorescence is separated in time and detected by a photodiode focused on the center of the light sheet, producing a time-of-flight trace. This detection method can also be used to characterize earlier stages of the experimental sequence. In this work, we analyze time-of-flight traces obtained after the double velocity selection stage and after a single Bragg diffraction pulse, in addition to the interferometry signals.
\subsection{\label{subsec:vs_theory}Velocity-selective Raman transitions}
A theoretical description of velocity-selective Raman transitions is given in \cite{vs_moler}. Consider a three-level \mbox{$\Lambda$-system} with two ground states $|g_1\rangle$ and $
|g_2\rangle$ split by energy $\hbar\omega_{21}$, both coupled to a single excited state $|e\rangle$, as illustrated in Fig.\,\ref{fig:vs_schematic}(a). Two counter-propagating light fields with angular frequencies $\omega_{1,L}$, $\omega_{2,L}$ and Rabi frequencies $\Omega_1$, $\Omega_2$ can transfer population from $|g_1\rangle$ to $|g_2\rangle$ with a probability $P_{12}(p)$ dependent on the momentum component $p$ parallel to the laser beams \cite{vs_moler}:
\begin{figure}
    \centering
    \includegraphics[width=1\linewidth]{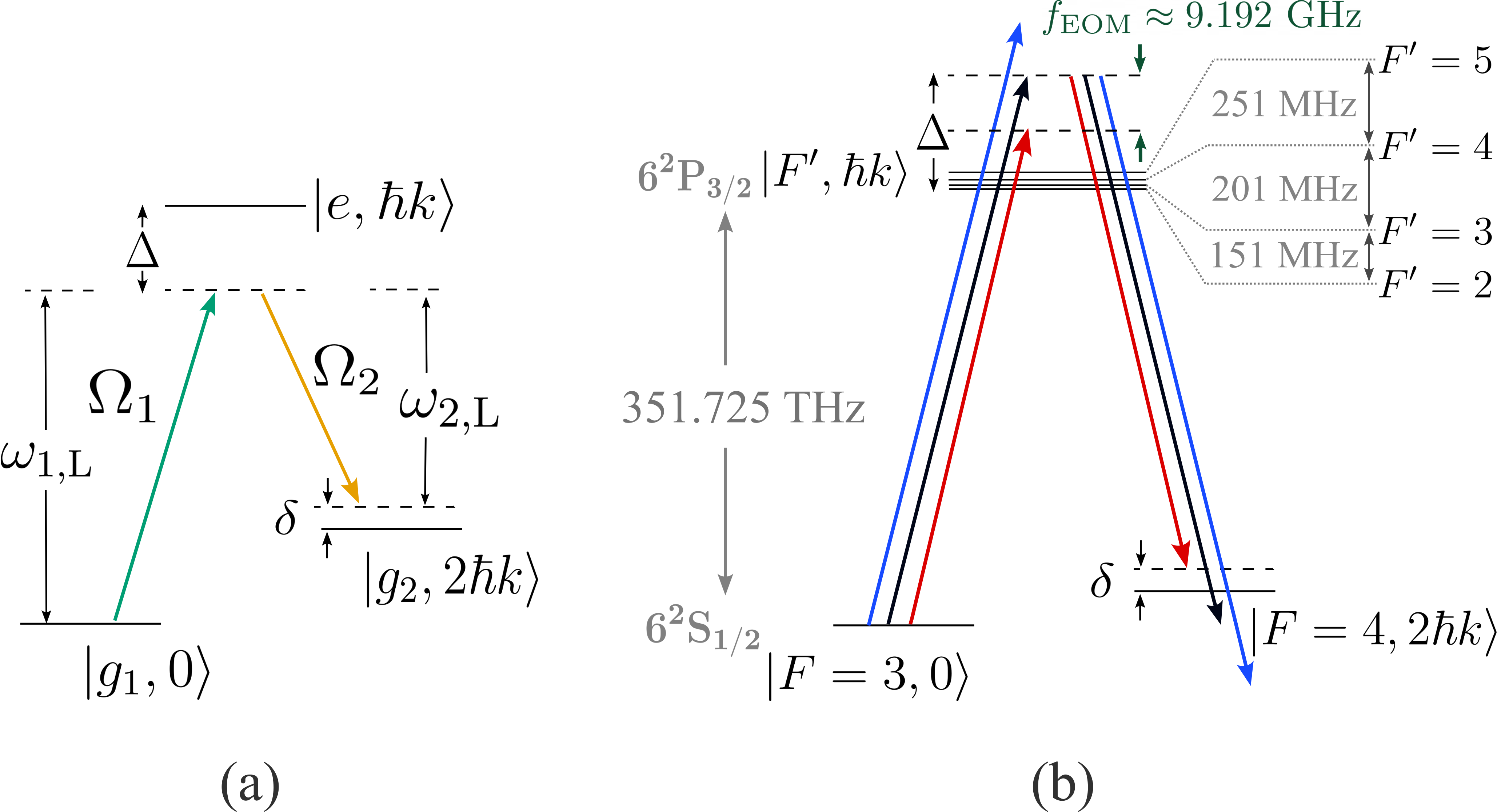}
    \caption{(a) Three-level Raman system and (b) the corresponding cesium level structure used in our experiment. Velocity-selective Raman transitions couple the $6^2S_{1/2}$ ground-state hyperfine levels of cesium via the $6^2P_{3/2}$ excited-state manifold.  A laser beam at frequency $f_0$ with single-photon detuning $\Delta$ from the $F=3\rightarrow F'=4$ transition is sent through an electro-optic modulator (EOM) driven with an RF frequency $f_{\mathrm{EOM}}$ near the hyperfine splitting frequency ($\approx 9.192~\mathrm{GHz}$), generating sidebands at frequencies $f_0\pm nf_{\mathrm{EOM}}, n=1,2,3,\ldots$. Here, we show only the carrier at $f_0$ and the first-order sidebands at $f_0 + f_{\mathrm{EOM}}$ and $f_0 - f_{\mathrm{EOM}}$, represented by black, blue, and red arrows, respectively. The EOM output propagates vertically into the vacuum chamber and is retro-reflected by a mirror above the interaction region, such that each frequency component is present in both propagation directions during the velocity selection pulse. The experiment operates with $\Delta$ in the range of $\numrange{5}{8}\,\mathrm{GHz}$, comparable in magnitude to the excited-state hyperfine splittings.}
    \label{fig:vs_schematic}
\end{figure}

\begin{equation}
\begin{split}
P_{12}&(p)=\\ \frac{1}{2} &\frac{\Omega_{\mathrm{eff}}^2}{\left(\delta(p) -\Delta_{\mathrm{AC}}\right)^2 + \Omega_{\mathrm{eff}}^2}\sin^2{\frac{\tau\sqrt{\left(\delta(p) -\Delta_{\mathrm{AC}}\right)^2 + \Omega_{\mathrm{eff}}^2}}{2}}
\end{split}
\label{eq:transfer_prob_simple}
\end{equation}
where we define the momentum-dependent detuning $\delta(p)$ as
\begin{equation}
\begin{split}
\delta(p) \equiv &-\frac{p\left(k_{1,L} + k_{2,L}\right)}{M} + \left(\frac{\left(\hbar k_{1,L}^2\right)}{2M} - \frac{\left(\hbar k^2_{2,L}\right)}{2M}\right)\\ &- \omega_{21} - \left(\omega_{2,L} - \omega_{1,L}\right),
\end{split}
\end{equation}
and define the quantities
\begin{equation}
\Delta_{\mathrm{AC}} \equiv \frac{|\Omega_1|^2}{\Delta} - \frac{|\Omega_2|^2}{\Delta},~ \Omega_{\mathrm{eff}}^2 \equiv 4\frac{|\Omega_1|^2|\Omega_2|^2}{\Delta^2}.
\label{eq:simple_deltaac_def}
\end{equation}
We refer to $\Delta_{\mathrm{AC}}$ as the differential light shift, and to $\Omega_{\mathrm{eff}}$ as the effective Rabi frequency.  $\Delta$ and $\tau$ denote the single-photon detuning and light pulse duration, respectively. Equation\,(\ref{eq:transfer_prob_simple}) is valid in the regime where $\Delta \gg |\Omega_1|, |\Omega_2|, \delta$, and after applying the rotating wave approximation \cite{vs_moler}. 

The differential light shift $\Delta_{\mathrm{AC}}$  always appears as a term subtracted directly from $\delta(p)$. Since $\mathrm{\Delta_{AC}}$ depends on $|\Omega_1|^2$ and $|\Omega_2|^2$, both of which scale linearly with the light intensity, the resonant velocity class is sensitive to the local laser intensity via the intensity dependence of $\mathrm{\Delta_{AC}}$.  Due to the Gaussian intensity profile of the Raman laser, we expect a radial variation of atomic velocities in the transferred population.

Our implementation of velocity selection introduces additional complexities, as illustrated in Fig.\,\ref{fig:vs_schematic}(b). First, the excited-state hyperfine levels splittings are not negligible compared to the single-photon detuning $\Delta$ (referenced to the $F=3\rightarrow F'=4$ transition), which is typically $\numrange{5}{8}\,\mathrm{GHz}$ in our experiment.
\begin{figure}[t]
    \centering
    \includegraphics[width=0.95\columnwidth]{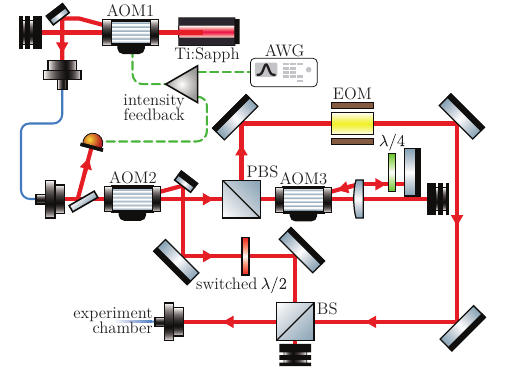}
    \caption{Schematic of the optical system used to generate the frequency pairs for velocity-selective Raman transitions. A Ti:Sapph laser is offset-locked to an external-cavity diode laser (not shown) stabilized to the Cs D2 $F=3\rightarrow F'=4$ transition. The offset frequency sets the Raman single-photon detuning, $\Delta$. The beam is sent through AOM1 for intensity stabilization, then split by AOM2, whose drive frequency controls the Raman two-photon detuning. One path is double-passed through AOM3, whose drive frequency is linearly ramped to compensate for the changing Doppler shift during free fall, then sent through an EOM driven near the ground state hyperfine splitting frequency. The EOM output is recombined with the other AOM2 path and coupled into a optical fiber leading to the vacuum chamber. The same optics chain is used for Bragg diffraction; in this mode, the EOM is disabled and the beam polarizations are switched to parallel using a voltage-controlled half-wave liquid  crystal retarder (``switched $\lambda/2$'').}
    \label{fig:ify_chain}
\end{figure}
Second, our optical system generates the required frequency components with an electro-optical modulator (EOM), which generates multiple sidebands. Figure \ref{fig:ify_chain} shows a simplified schematic of the optics chain used in the Raman laser system. All optical frequency components are present during the Raman pulse in both the upward- and downward-propagating directions. Thus, multiple Raman pathways can contribute to the population transfer, and the differential light shift must be calculated by accounting for all relevant frequency components and excited-state hyperfine levels \cite{Wu_2017}. Therefore, accounting for the effects neglected in the idealized expression for $\Delta_{\mathrm{AC}}$ given in Eq.\,(\ref{eq:simple_deltaac_def}), a more accurate expression is 
\begin{equation}
\Delta_{\mathrm{AC}}\equiv \sum_{s=-\infty}^{\infty}\sum_{F', m_F',q}\left(\frac{|\Omega^{(s,q)}_{3,0\rightarrow F',m_F'}|^2}{4\Delta^{(s)}_{3,F'}} - \frac{|\Omega^{(s,q)}_{4,0\rightarrow F',m_F'}|^2}{4\Delta^{(s)}_{4, F'}}\right),
\label{eq:magic_condition}
\end{equation}
where $n$ enumerates the EOM sidebands and $F'$ indexes the hyperfine levels in the excited-state manifold. The relative optical power in the $s$-th EOM sideband is $J^2_s\left(\beta\right)$, where $J_s$ is the $s$-th order Bessel function of the first kind and $\beta$ is the phase modulation index. The index $q = -1, 0, +1$ labels the spherical polarization components ($\sigma^-$, $\pi$, and $\sigma^+$, respectively) of the light fields with respect to the quantization axis defined by the magnetic field, which is oriented along the launch direction of the atoms. In our experiment, the Raman beams propagate parallel to the quantization axis and are linearly polarized perpendicular to it. Consequently, each beam ideally consists of an equal superposition $q=\pm 1$ components and no $q=0$ component. In practice, small polarization or alignment imperfections can introduce a weak $q=0$ component, which is included in Eq.\,(\ref{eq:magic_condition}).

We define the \textit{magic detuning} as the value of the single-photon detuning $\Delta$ where $\Delta_{\mathrm{AC}} = 0$, making the resonant velocity condition invariant to the laser intensity. In principle, the magic detuning can be calculated from Eq.\,(\ref{eq:magic_condition}); however, the result depends on the relative intensities of all optical frequency components, which are difficult to measure accurately.

\subsection{\label{subsec:bragg_diffraction}Phase Contributions from Bragg Diffraction}
\begin{figure}
    \centering
     \includegraphics[width=0.92\linewidth]{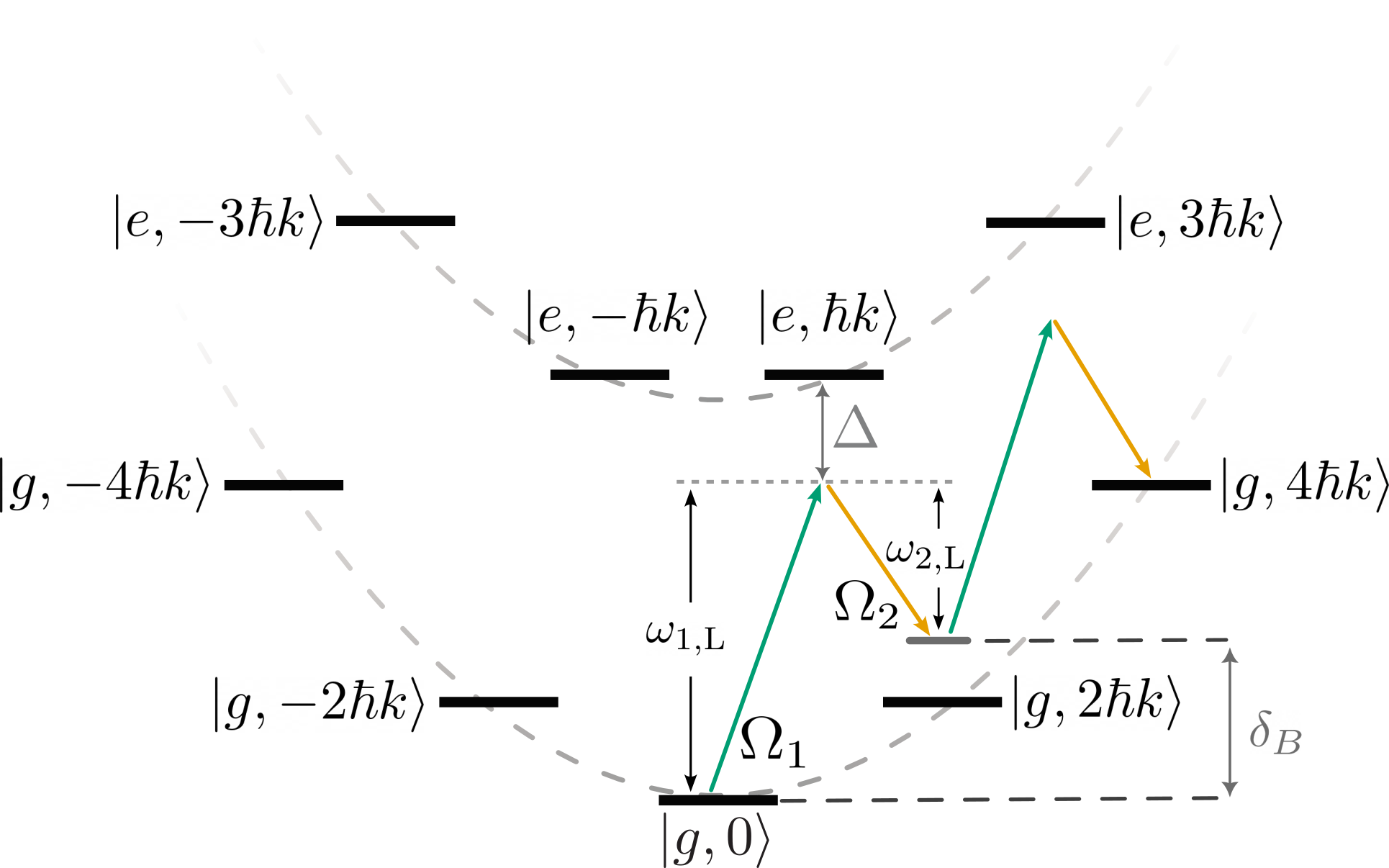}
    \caption{Schematic of Bragg diffraction dynamics $(n=2)$. Counter-propagating laser beams with angular frequencies $\omega_{\mathrm{1,L}}$ and $\omega_{\mathrm{2,L}}$, and frequency difference $\delta_B\equiv\omega_{\mathrm{1,L}}-\omega_{\mathrm{2,L}}$ are used to coherently couple momentum states spaced by integer multiples of $2\hbar k$. $\Omega_1$ and $\Omega_2$ denote the single-photon Rabi frequencies.}
    \label{fig:bragg_level_structure}
\end{figure}

\begin{figure}
    \centering
    \includegraphics[width=1.0\linewidth]{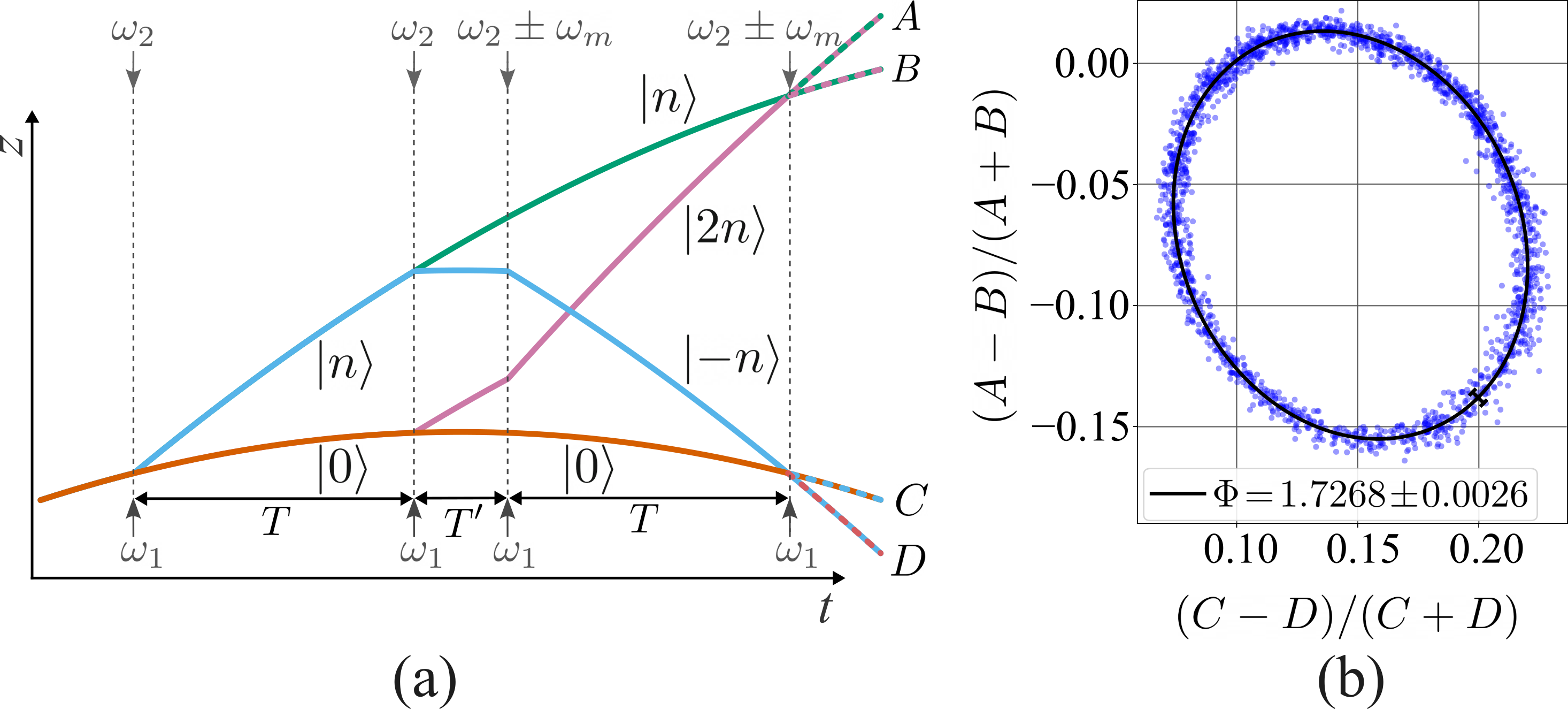}
    \caption{(a) Geometry of a simultaneous conjugate Ramsey-Bord\'e interferometer. Four Bragg pulses (dashed lines) split and recombine an atomic ensemble into two conjugate interferometers. The first two pulses create four distinct momentum components, while the final two pulses, driven by frequencies, $\omega_1$ and $\omega_2\pm\omega_m$, simultaneously address both interferometers.
We measure the differential phase between the two conjugate interferometers to reject common-mode noise sources such as vibrations. Output populations are labeled $A$-$D$. (b) Sample data (2000 shots) from the interferometry sequence using 4th order Bragg pulses ($T=\mathrm{50~ms}$, $\mathrm{\Delta = 6675~MHz}$, $57\,\mathrm{mW}$ Raman power).  Blue points represent individual experimental shots; the abscissas and ordinates correspond to $\left(C-D\right)/\left(C+D\right)$ and $\left(A-B\right)/\left(A+B\right)$, respectively. The black curve is the Bayesian ellipse fit used to extract the differential phase $\Phi$ and its uncertainty.}
    \label{fig:ify_geometry}
\end{figure}

Bragg diffraction coherently couples momentum states separated by integer multiples of $2\hbar k$ through a virtual excited state. A moving optical lattice is formed by counter-propagating laser beams with angular frequencies $\omega_{\mathrm{1,L}}$ and $\omega_{\mathrm{2,L}}$, whose frequency difference we denote as $\delta_B \equiv \omega_{\mathrm{1,L}} -\omega_{\mathrm{2,L}}$. The beams are detuned from the excited state by a large single-photon detuning $\Delta$ ($\delta_B \ll \Delta$). An $n$-th order Bragg process transfers $2n\hbar k$ of momentum and is resonant when
\begin{equation}
    \delta_B = 4n\omega_r + 2kv,
    \label{eq:delta_B}
\end{equation}
where $\omega_r = \hbar  k^2/2m$ is the recoil frequency and $kv$ is the Doppler shift for an atom moving at velocity $v$ parallel to the laser beams (see Fig.\,\ref{fig:bragg_level_structure} and Ref. \cite{holger_bragg}).
Off-resonant couplings to intermediate momentum states imprint additional phase shifts on the atomic wavefunctions during Bragg diffraction. This diffraction phase contributes an additional term, $\phi_D$, to the measured interferometer phase, $\Phi$, in the simultaneous conjugate Ramsey–Bordé interferometer geometry (Fig.\,\ref{fig:ify_geometry}(a)):

\begin{equation}\Phi = 16n\left(n+N\right) \omega_r T - 2n\omega_m T + \phi_D, 
\label{eq:ideal_equation}
\end{equation}
where $T$ is the separation between the first and second pulses (and between the third and fourth pulses).  $\phi_D$ can reach $\mathord{\sim}100~\mathrm{milliradians}$, making it a significant systematic effect. The four output populations of the interferometer, labeled $A$-$D$ in Fig.\,\ref{fig:ify_geometry}(a), are used to extract the interferometer phase. In our interferometer, the common-mode phase of the two simultaneuous conjugate interferometers varies randomly shot-to-shot due to vibration noise.  As a result, when the quantities $\left(C-D\right)/(C+D)$ and $(A-B)/(A+B)$ from each experimental shot are plotted as the abscissas and ordinates of a scatter plot, the data trace out an ellipse. By fitting the ellipse using a Bayesian estimator, one extracts $\Phi$ and its associated uncertainty. Figure \ref{fig:ify_geometry}(b) shows sample data consisting of 2000 experimental shots taken under a single set of operating conditions.

Our apparatus is designed for a precision measurement of the fine structure constant through the cesium recoil frequency \cite{berkeley_alpha}. In the full measurement configuration, we perform $N$ Bloch oscillations between the second and third Bragg pulses to increase the accumulated phase. Typical operating parameters  ($n=\numrange{4}{6}$, $N=\numrange{80}{100}$, and  $T\sim100~\mathrm{ms}$) result in a phase accumulation of $\mathord{\sim} 10~\mathrm{megaradians}$ for $852~\mathrm{nm}$ laser light. Thus, $\phi_D$ must be controlled or characterized at the milliradian level to achieve sub-ppb accuracy in the recoil measurement. (The present work focuses exclusively on systematic phase shifts arising from Bragg diffraction and does not include Bloch oscillations).

\begin{figure}
    \centering
    \includegraphics[width=0.82\linewidth]{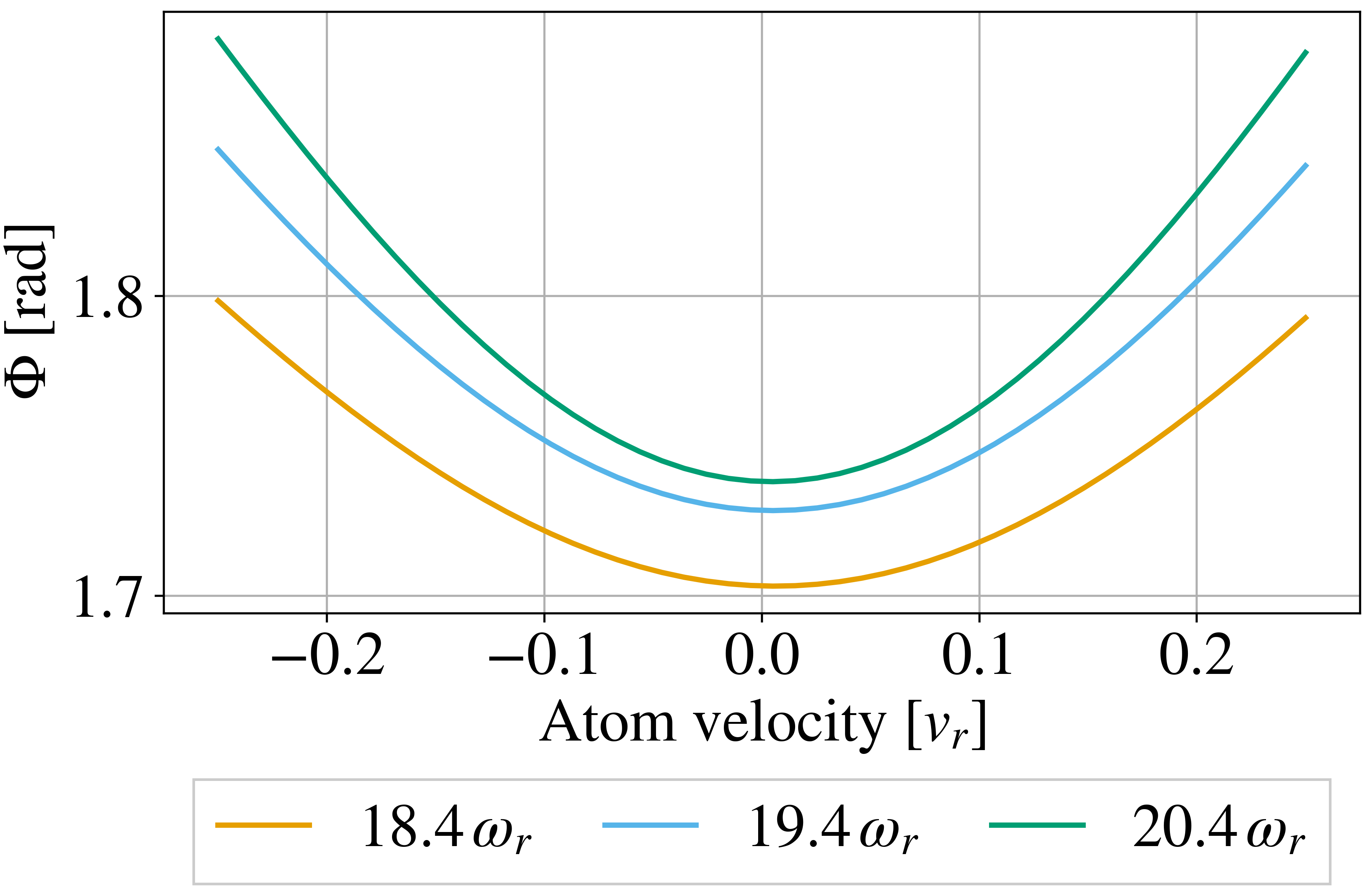}
    \caption{Simulated interferometer phase $\Phi$ as a function of atom velocity in units of cesium recoil velocities, $v_r$, computed by numerically propagating single-atom wavefunctions through the full interferometery sequence for $n=4$ and $T=50~\mathrm{ms}$. The different colored curves indicate different values of $\Omega_{B,\mathrm{eff}}$, the local Bragg Rabi frequency in units of cesium recoil frequencies, $\omega_r$.}
    \label{fig:diffraction_phase}
\end{figure}

 The diffraction phase has an undesired velocity dependence due to the Doppler dependence of the two-photon resonance condition \cite{estey_2015}. To illustrate this effect, we numerically propagate single-atom wavefunctions through the full interferometry sequence for $n=4$, $N=0$, and $T=\mathrm{50~ms}$. Figure \ref{fig:diffraction_phase} shows the interferometer phase as a function of atomic velocity with $\delta_B= 16\omega_r$, corresponding to the resonance condition for zero-velocity atoms undergoing 4th-order Bragg diffraction. As the velocity changes, the Doppler shift detunes the Bragg transition and produces a velocity-dependent diffraction phase with a minimum at resonance. This feature can used to identify the resonant value of $\delta_B$ \cite{estey_2015}. The diffraction phase also varies with the local Bragg beam intensity through the effective Rabi frequency, $\Omega_{B,\mathrm{eff}}\equiv \Omega_1\Omega_2/\Delta$, where $\Omega_1$ and $\Omega_2$ are the single-photon Rabi frequencies; therefore, we compute these curves for three different values of $\Omega_{B,\mathrm{eff}}$. 

\section{\label{sec:methods} Experimental Methods}
The diffraction phase depends on the atomic velocity, while differential light shifts locally modify the resonant velocity class during velocity selection. Consequently, varying the intensity of the velocity-selective Raman pulse while holding all other experimental parameters fixed is expected to change the ensemble-averaged diffraction phase. Our goal is to experimentally measure the dependence of the diffraction phase on the Raman pulse intensity. Additionally, we investigate whether the sensitivity of the diffraction phase to Raman pulse intensity can be suppressed by operating at the Raman single-photon detuning where $\mathrm{\Delta_{AC}} = 0$, which eliminates the spatial inhomogeneity of the atomic velocity distribution.

Prior to interferometer operation, $\delta_B$ and $\Omega_{B,\mathrm{eff}}$ are first coarsely optimized at the point of equal output population between the diffracted and undiffracted momentum states after a single Bragg pulse, inferred from a time-of-flight trace. To determine the resonant value of $\delta_B$ to sub-kHz precision, we run the full interferometer sequence at several values of $\delta_B$ spanning a range of $\mathrm{1\text{-}2~{kHz}}$ around the initial estimate of the resonant frequency difference. 

Repeating this procedure for multiple values of $\delta_B$ produces parabola-shaped curves (Fig.\,\ref{fig:testplot}). Interferometer parameters were fixed at $T=50~\mathrm{ms}$, $T' = 8~\mathrm{ms}$, and $n=4$. The resonant value of $\delta_B$ and the corresponding differential phase are determined by performing a quadratic fit to the data points and extracting the coordinates of the parabola's vertex.

We performed the procedure described above for a range of single-photon Raman detunings $\Delta$ and Raman powers. $\Delta$ was adjusted over a range of $\numrange{5}{8}\,\mathrm{GHz}$ by changing the Ti:Sapph frequency (Fig.\,\ref{fig:ify_chain}). The intensities of both the Raman and Bragg pulses are stabilized using an intensity servo with a bandwidth of approximately $\mathrm{1~MHz}$. The servo monitors a small fraction of the light diffracted by AOM1 with a photodiode and actively adjusts the AOM1 RF drive power to match the photodiode signal to a reference waveform generated by an arbitrary waveform generator. The Raman pulses consist of   $\mathrm{400~\mu s}$ square pulses with short ramps at the leading and trailing edges, while the Bragg pulses consist of Gaussian temporal profiles with a $1\sigma$ width of $\mathrm{20~\mu s}$. The optical power calibration is approximately $0.6~\mathrm{mW}/\mathrm{mV}$, for a $1/e^2$ beam waist of $\mathrm{6.2~mm}$. For both Raman and Bragg, the RF drive power to AOM2 is set so that the diffracted and undiffracted beams have approximately equal intensities.

\begin{figure}
    \centering
    \includegraphics[width=0.95\linewidth]{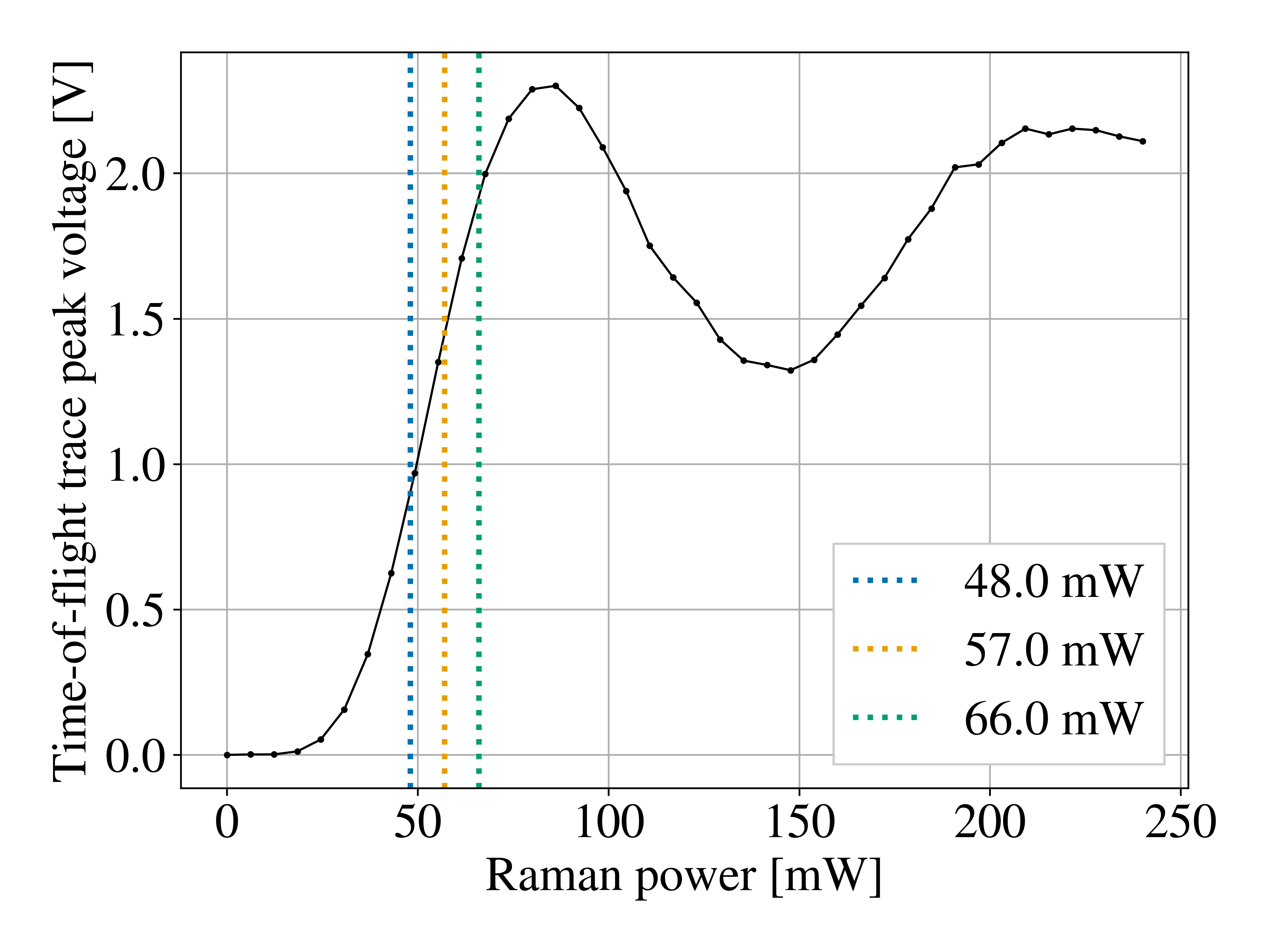}
    \caption{Sample Rabi flopping curve used to select operating points for Raman powers at $\Delta = \mathrm{7475~MHz}$. The black curve is comprised of the peak amplitudes extracted from Gaussian fits to time-of-flight traces acquired after double velocity selection over a range of Raman powers. The three colored dashed lines indicate the three Raman powers, below the peak of the first Rabi oscillation, that were chosen as operating points for experiments at this value of $\Delta$. The reduced contrast of the Rabi oscillations is expected due to spatial inhomogeneity of the Raman beam: near the peak of the first oscillation, atoms in higher-intensity regions of the beam may already be evolving downward along the Rabi flop, while atoms in lower-intensity regions are still evolving upward.}
    \label{fig:rabiflop}
\end{figure}

The operating points for the Raman powers were chosen as follows. For a given value of $\Delta$, we first performed the standard state preparation sequence concluding with double velocity selection and recorded a time-of-flight trace. The trace was fitted to a Gaussian profile, and the fitted amplitude was used as an estimate of a quantity proportional to the population transferred within the resonant velocity class. Repeating this procedure over a range of Raman powers produces a Rabi-flopping curve, as shown in Fig.\,\ref{fig:rabiflop}. For each value of $\Delta$, the Raman power maximizing population transfer was identified from the first Rabi oscillation. Three Raman powers spanning a range of $18\,\mathrm{mW}$ below this nominal $\pi$-pulse amplitude were selected for each value of $\Delta$.

In our current experimental configuration, the same laser system and optics are used for both velocity-selective Raman transitions and Bragg diffraction. Consequently, the Bragg single-photon detuning varies with the Raman single-photon detuning, requiring the Bragg pulse power to be adjusted to maintain a constant $\Omega_{B,\mathrm{eff}}$. For each value of $\Delta$, a single-frequency Bragg power (for the first and second pulses) and a corresponding dual-frequency Bragg power (for the third and fourth pulses) were selected and used for all three Raman powers (see Fig.\,\ref{fig:ify_geometry}). The Bragg powers were determined using the largest of the three Raman powers selected for each value of $\Delta$. After performing state preparation through double velocity selection, we applied a single-frequency Bragg pulse and recorded the resulting time-of-flight signal. This measurement was repeated over a coarse two-dimensional scan of single-frequency Bragg power and $\delta_B$. For each scan point, we calculated the population inversion, $(p_1-p_2)/(p_1+p_2)$, where $p_1$  and $p_2$ are the diffracted and undiffracted atomic populations extracted from the time-of-flight signal. The single-frequency Bragg power was chosen such that the population inversion was closest to the midpoint between its minimum and maximum observed values. The corresponding dual-frequency power was then set to twice this value. Repeating the optimization for the remaining two Raman powers produced no measurable change in the optimal Bragg power.

$\Delta_{\mathrm{AC}}$ is proportional to the Raman power, while $\Delta$ controls the strength of this dependence. Consequently, as $\Delta$ is moved farther from the magic detuning, a fixed change in Raman power is expected to produce a larger shift in the mean velocity of the selected atoms, leading to a correspondingly larger shift in the value of $\delta_B$ that minimizes the diffraction phase. In addition to changing the mean velocity, varying the Raman power affects the spatial variation of the selected velocity class. Because the sensitivity of this spatial variation to the Raman power increases with distance from the magic detuning, we expect the same range of Raman powers ($18\,\mathrm{mW}$) to produce increasingly distinct ensemble-averaged diffraction phases at the resonant value of $\delta_B$ as $\Delta$ is moved away from the magic detuning. Near the magic detuning, where $\mathrm{\Delta_{AC} = 0}$, we expect both effects to be suppressed. 
\section{\label{sec:results} Results and Discussion}
\renewcommand{\arraystretch}{1.2}
\begin{table}
\centering
\begin{tabular}{|c |c|c|c|}
\hline
$\Delta$ [MHz] & Power [mW]& Resonant $\delta_B$ [kHz]& $\Phi$ min. [rad] \\
\hline
 & 36 & $230.05\pm0.02$ & $1.445 \pm 0.002$\\
5300 & 45 & $230.20\pm0.02$ & $1.447\pm0.002$ \\
 & 54 & $230.34\pm0.02$ & $1.448 \pm 0.002$\\
 \cline{1-4}
 & 45 & $229.73\pm0.01$ & $2.656\pm0.001$ \\
5675 & 57 & $229.78\pm0.01$ & $2.654\pm0.001$ \\
 & 63 & $229.84\pm0.01$ & 
 $2.657\pm0.001$\\
 \cline{1-4}
 & 48 & $225.18\pm0.03$ & $0.973\pm0.002$ \\
7475 & 57 & $224.33\pm0.04$ & $0.983\pm0.002$ \\
 & 66 & $223.35\pm0.03$ & $0.988\pm0.002$ \\

\hline
\end{tabular}
\caption{Minimum $\Phi$ and corresponding resonant $\delta_B$-values for each combination of $\Delta$ and Raman power, extracted from parabolic fits to the experimental data in Figure \ref{fig:testplot}.}
\label{tab:fitted_minima}
\end{table}
  Figures \ref{fig:testplot}(a)-(c) show the diffraction phase landscape, i.e., the dependence of the diffraction phase on $\delta_B$, for three representative values of $\Delta$. Figure \ref{fig:testplot}(b) shows the results for $\mathrm{\Delta=5675~MHz}$, which we determined to be the closest tested value of $\Delta$ to the magic detuning condition. The fitted parabolic curves obtained at the three Raman pulse powers nearly overlap, indicating a negligible dependence on Raman power. In contrast, Figs. \ref{fig:testplot}(a) and \ref{fig:testplot}(c) show the results for $\mathrm{\Delta = 5300~MHz}$ and $\mathrm{\Delta = 7475~MHz}$, respectively, corresponding to single-photon detunings below and above the magic condition. In these cases, the parabola vertices no longer coincide horizontally. This behavior is expected: in the presence of a differential light shift proportional to the Raman power, the resonant velocity condition is modified. As a result, the average velocity of the atomic cloud changes, shifting the Bragg resonance condition.  We also note that for $\mathrm{\Delta=5300~MHz}$, the resonant value of $\delta_B$ increases as the Raman power is increased, while the opposite trend is observed for $\mathrm{\Delta = 7475~MHz}$. This indicates that $\mathrm{\Delta_{AC}}$ has opposite signs for $\mathrm{5300~MHz}$ and $\mathrm{7475~MHz}$.

In Table \ref{tab:fitted_minima}, we summarize the fitted minimum phase and corresponding resonant $\delta_B$-values for each combination of $\Delta$ and Raman power. Notably, for $\mathrm{\Delta=7475~MHz}$, where the magnitude of the differential light shift is relatively large (see Section \ref{sec:mwl_validation}, Fig.\,\ref{fig:ac_stark_measurement}), the minimum diffraction phases given by the ordinates of the fitted parabola vertices vary by more than $\mathrm{10~mrad}$ across the range of Raman powers studied. This observation implies that applying the standard diffraction phase minimization procedure described in the previous section would yield different optimal diffraction phases depending on the Raman power used during the state-preparation stage. The observed intensity dependence of the minimum diffraction phase demonstrates that operating away from the magic velocity selection condition introduces an additional systematic sensitivity to the Raman power. Consequently, fluctuations or drifts in the state-preparation laser intensity can change the optimized diffraction phase and thereby contribute to the interferometer phase uncertainty.

In contrast, this sensitivity is suppressed near the magic detuning. These results have important implications for numerical modeling. Away from the magic detuning, the light shift during velocity-selective Raman transitions couples the state-preparation dynamics to the subsequent Bragg diffraction phase by generating a spatial dependence in the selected atomic velocity distribution. Accurate prediction of the diffraction phase therefore requires detailed knowledge of the spatial correlations between atomic position, velocity, and Raman beam intensity profile during state preparation. Near the magic detuning, these correlations are largely eliminated because the differential light shift is suppressed. Operating at the magic detuning therefore not only reduces systematic sensitivity to Raman laser intensity fluctuations but also substantially simplifies numerical modeling of the diffraction phase. 
\section{\label{sec:mwl_validation} Additional Experimental Validation of the Magic detuning}
We used two approaches to confirm the location of the velocity selection magic detuning in our experimental setup: 1) measuring the time-of-flight peak arrival time after double velocity selection and 2) tracking the Bragg resonance condition from a time-of-flight trace obtained after a single Bragg pulse.
\subsection{\label{subsec:arrival_time_validation} Velocity selection time-of-flight traces}
\begin{figure}
    \centering
    \includegraphics[width=0.95\linewidth]{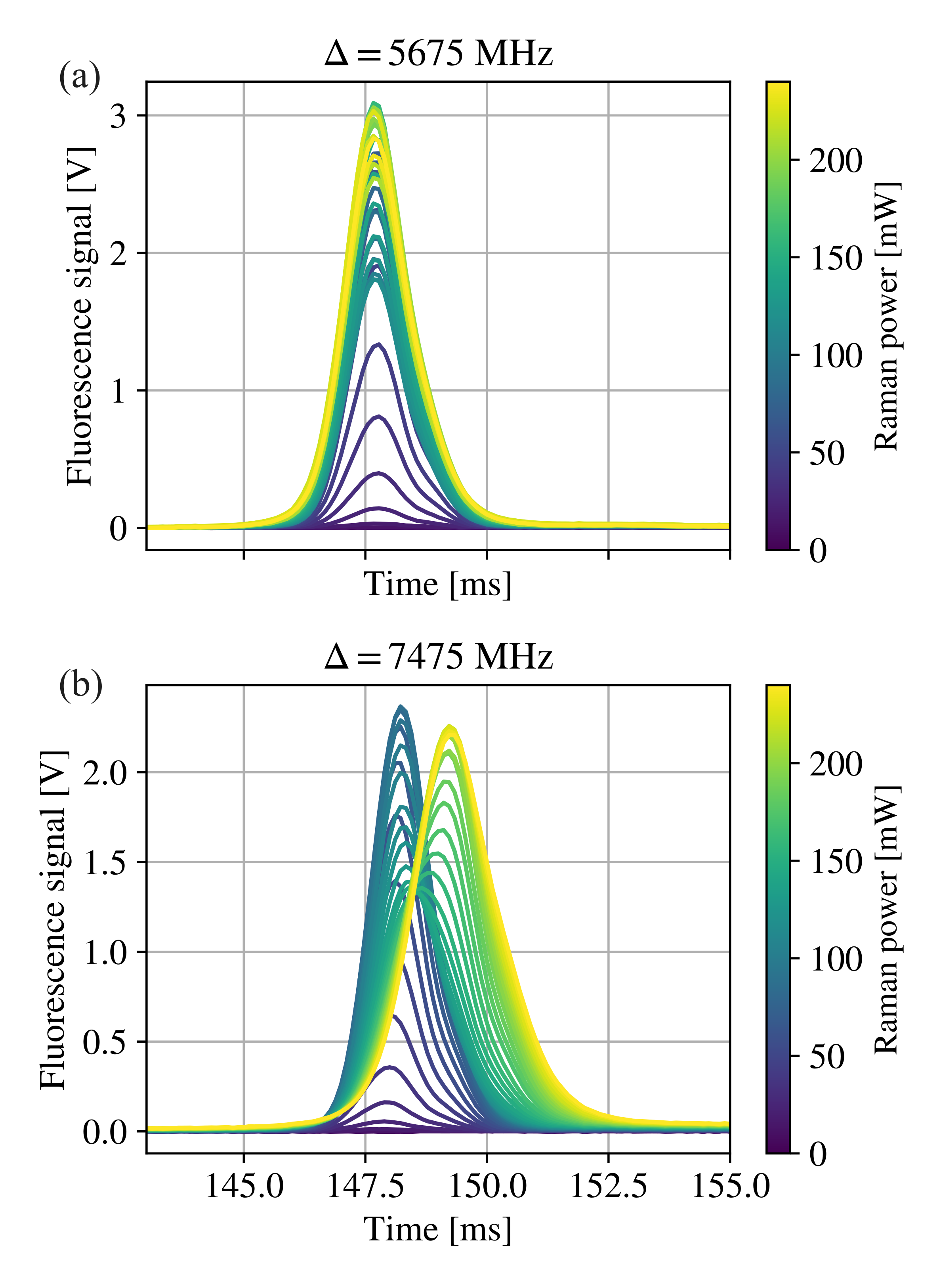}
    \caption{Time-of-flight traces after double velocity selection as Raman power is varied for (a) $\Delta = \mathrm{5675~MHz}$ (near magic detuning) and (b) $\Delta = \mathrm{7475~MHz}$ (away from the magic detuning). In (a), the peak height varies as Raman power is increased due to Rabi oscillations, but the arrival time of the peak remains invariant. Meanwhile, in (b), both the peak height and arrival time vary as the Raman power is varied, indicating a non-zero $\mathrm{\Delta_{AC}}$.}
    \label{fig:arrival_time_example} 
\end{figure}

\begin{figure}
    \centering
    \includegraphics[width=0.95\linewidth]{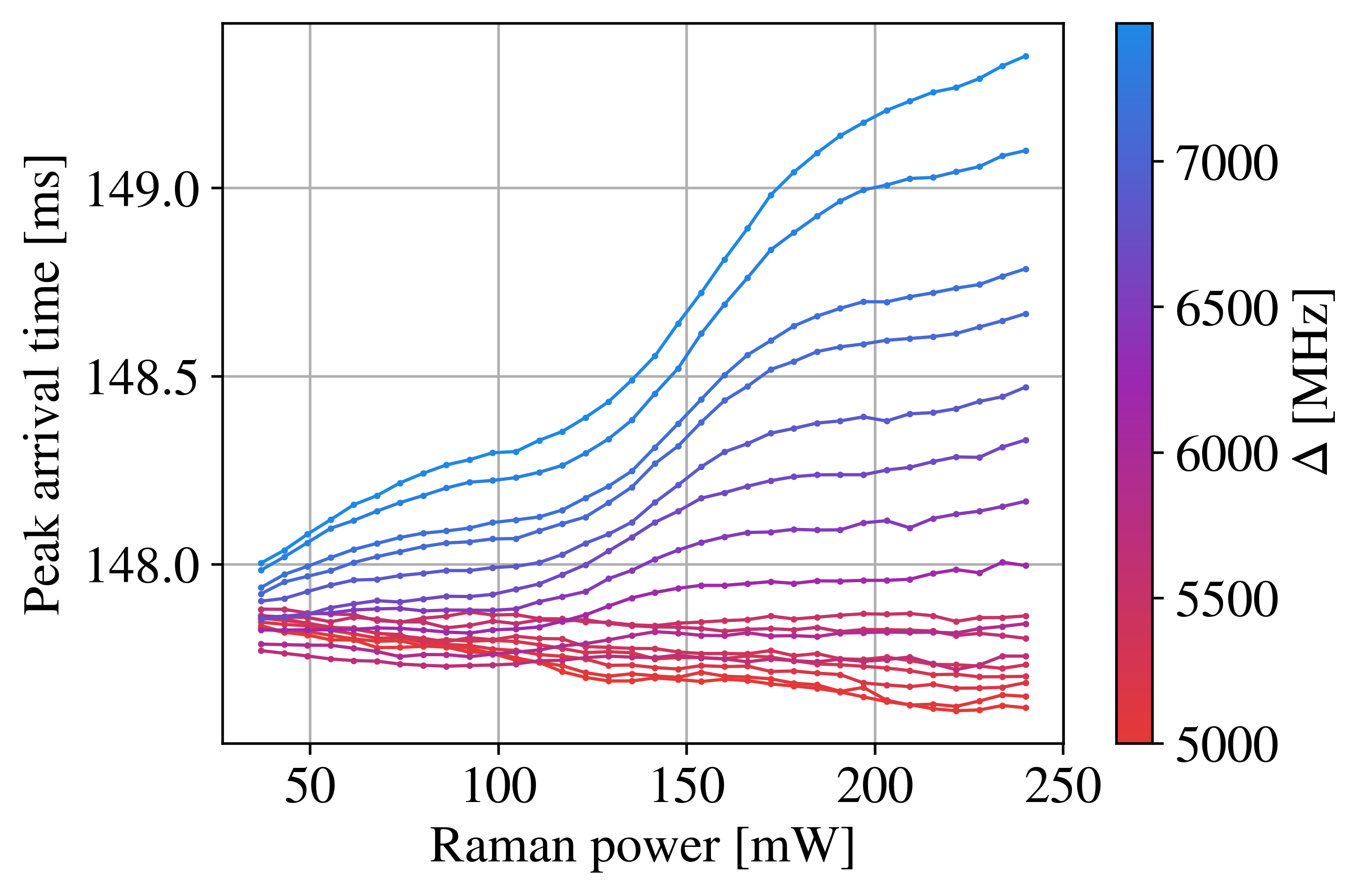}
    \caption{Velocity selection time-of-flight peak arrival times as a function of Raman power, for a range of values of $\Delta$. The magic detuning corresponds to a flat slope.}
    \label{fig:arrival_time_summary}
\end{figure}

The peak of a time-of-flight trace after double velocity selection corresponds to atoms in the velocity class that is transferred with highest efficiency. Thus, away from the magic detuning, the peak arrival time is expected to vary with Raman power because the resonant velocity condition changes with the magnitude of $\mathrm{\Delta_{AC}}$. We overlay the time-of-flight traces recorded over a range of Raman powers in Figure \ref{fig:arrival_time_example}(a) (at $\Delta = \mathrm{5675~MHz}$, near the magic detuning) and Figure \ref{fig:arrival_time_example}(b) (at $\Delta=\mathrm{7475~MHz}$, far from the magic detuning). At both detunings, the peak amplitude of the time-of-flight traces varies with Raman power as the atoms undergo Rabi oscillations. However, in the far-from-magic-detuning plot, the arrival time of the time-of-flight peak also shifts as the Raman power is varied. We repeat this procedure for a range of $\Delta$ values. Each time-of-flight trace is fit to a Gaussian temporal profile, and the fitted mean is plotted as a function of Raman power for each value of $\Delta$. The results are shown in Fig.\,\ref{fig:arrival_time_summary}. For values of $\Delta$ blue-detuned with respect to $\mathrm{5675~MHz}$, the peak arrives progressively earlier in time as Raman power is increased. In contrast, for  values of $\Delta$ red-detuned with respect to $\mathrm{5675~MHz}$, the slope reverses sign and the time-of-flight peak shifts later in time with increasing Raman power. This behavior is consistent with a change in the sign of $\mathrm{\Delta_{AC}}$. However, due to the finite bandwidth of our detection system, this method is unable to resolve the magic detuning more precisely than a few hundred $\mathrm{MHz}$; from Fig.\,\ref{fig:arrival_time_summary}, we infer that the magic detuning is within the range $\numrange{5200}{6000}\,\mathrm{MHz}$.
\subsection{\label{subsec:arrival_time_validation} Resonance curves from single Bragg pulse diffraction efficiency}
\begin{figure}
    \centering
    \includegraphics[width=0.8\linewidth]{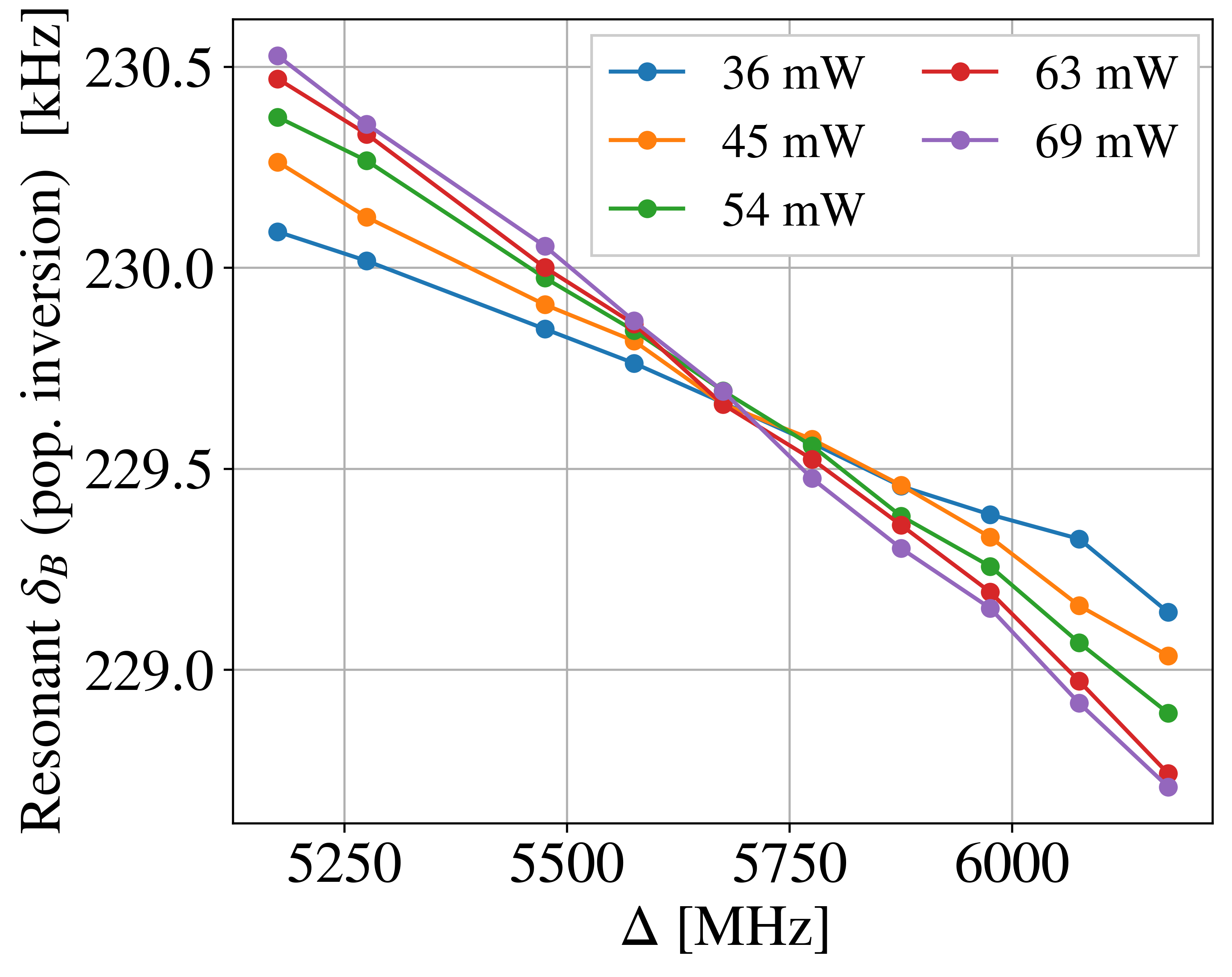}
    \caption{Resonant $\delta_B$-values determined from single Bragg pulses as a function of $\Delta$ for several Raman powers, indicated by the different colored curves. The common intersection of the curves identifies the magic detuning, which we estimate to be $\mathrm{5675\pm10~MHz}$.} 
    \label{fig:bragg_intersection_curve}
\end{figure}

An alternative method for determining the velocity selection magic detuning uses single Bragg diffraction pulses instead of the full interferometer sequence. For a given value of $\Delta$, we fix the Raman power at $\mathrm{69~mW}$ and choose a Bragg power using the coarse two-dimensional optimization procedure described in Section \ref{sec:methods}. Next, with the Bragg power held fixed, we measure the population inversion as a function of $\delta_B$ for several values of Raman power.
\begin{figure}
    \centering
    \includegraphics[width=0.9\linewidth]{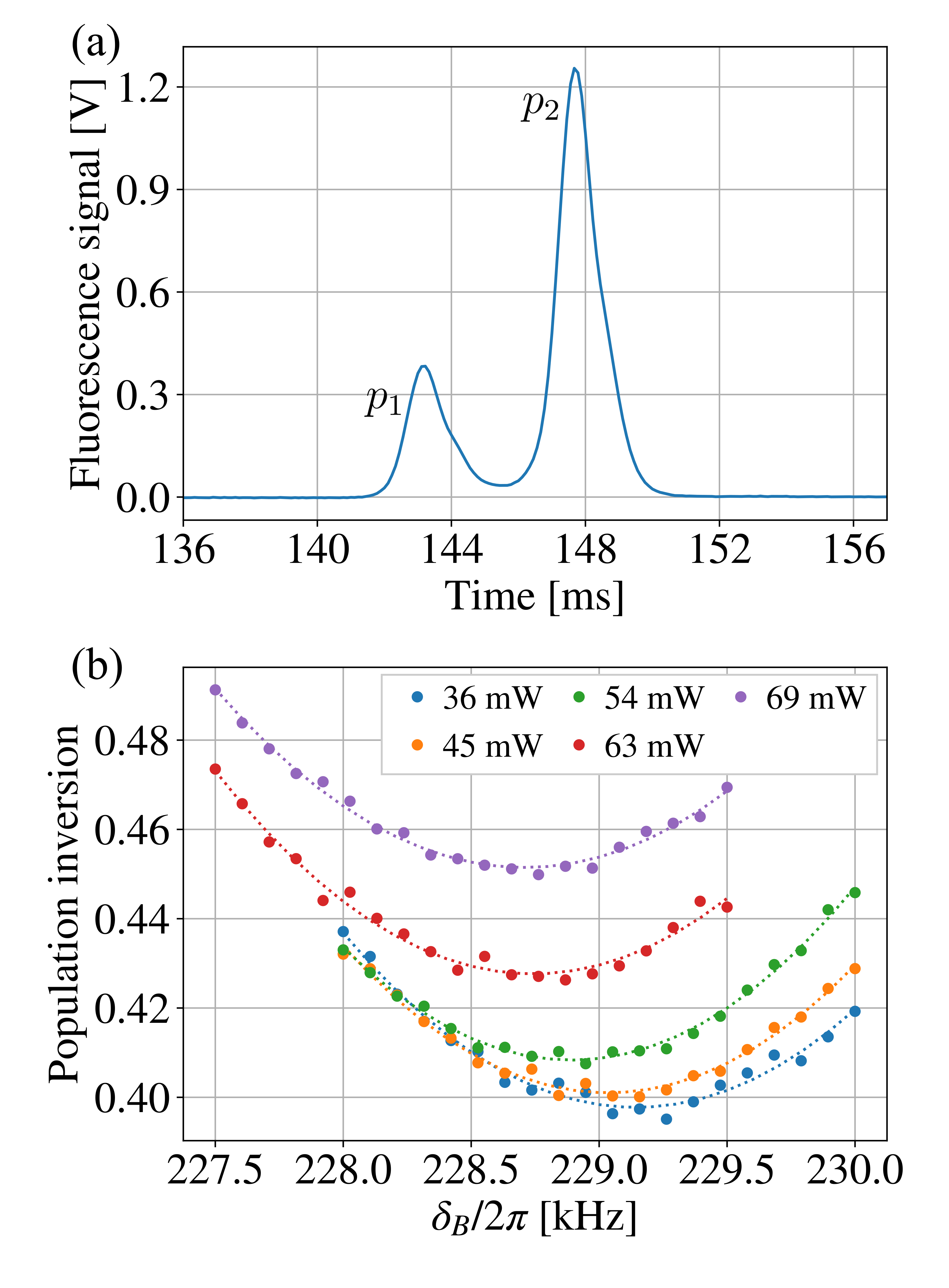}
    \caption{(a) Time-of-flight trace following a single $n=4$ Bragg diffraction pulse, from which the diffracted ($p_1$) and undiffracted ($p_2$) populations are extracted. (b) Population inversion, $(p_1-p_2)/(p_1+p_2)$, as a function of $\delta_B$ for several Raman powers with $\mathrm{\Delta=6175~MHz}$. Quadratic fits to the parabola-shaped curves are used to extract the resonant value of $\delta_B$ corresponding to the $\pi /2$ Bragg pulse condition.} 
    \label{fig:bragg_mwl_example}
\end{figure}
Each scan of $\delta_B$, spanning $\numrange{1}{2}\,\mathrm{kHz}$, produces a parabola-like curve, whose minimum corresponds to the resonant value of $\delta_B$ for a $\frac{\pi}{2}$-pulse. We fit each curve to a quadratic function to extract the value of $\delta_B$ where the inversion is minimized. The fitted resonant detunings as a function of $\Delta$ are plotted in Fig.\,\ref{fig:bragg_intersection_curve}.

The intersection of the $\Delta$-versus-resonant-$\delta_B$ curves measured at different Raman powers identifies the magic detuning, because this is the point at which the resonant value of $\delta_B$ is independent of the intensity of the preceding velocity selection pulses. From Eq.\,(\ref{eq:delta_B}), this implies that the mean velocity of the selected ensemble is invariant to the Raman pulse intensity. We estimate the intersection of the curves Fig.\,\ref{fig:bragg_intersection_curve} to be $\mathrm{5675\pm 10~MHz}$, consistent with the expected location of the magic detuning inferred from our diffraction phase studies with the full interferometer sequence as well as with velocity selection time-of-flight peak arrival time method.

\begin{figure}
    \centering
    \includegraphics[width=0.87\linewidth]{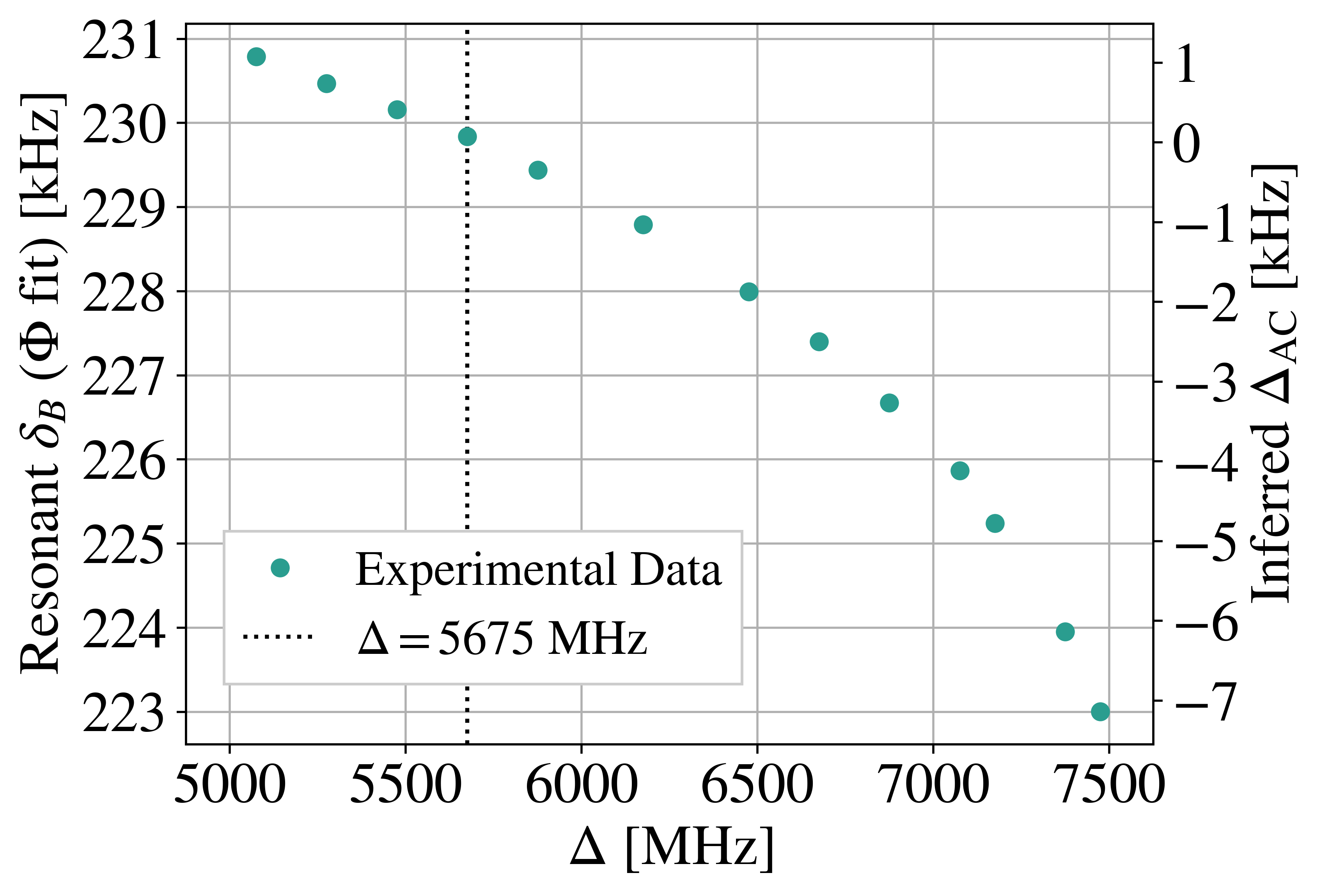}
    \caption{Left vertical axis: resonant $\delta_B$-values determined from diffraction phase measurements as a function of $\Delta$, with Raman power fixed at $\mathrm{69 ~mW}$. Right vertical axis: inferred $\Delta_{\mathrm{AC}}$ at this Raman power, assuming a magic detuning of $\Delta=\mathrm{5675~MHz}$.}
    \label{fig:ac_stark_measurement}
\end{figure}

Identifying the magic detuning, combined with measurements of the resonant values of $\delta_B$ obtained through diffraction phase measurements  (i.e. using the full interferometer sequence), enables an indirect measurement of $\mathrm{\Delta_{AC}}$ in kHz. We recorded diffraction phase data for several values of $\Delta$ while keeping Raman power fixed at $\mathrm{69~mW}$. The data in Fig.\,\ref{fig:ac_stark_measurement} correspond to the fitted resonant value of $\delta_B$ for each value of $\Delta$. By subtracting the inferred value of $\delta_{B}$ at the identified magic wavelength, we estimate the value of $\Delta_{\mathrm{AC}}$ during the velocity-selective Raman pulses as a function of $\Delta$. 
\section{\label{sec:conc}Conclusion and Outlook}
In this work, we have investigated the effect of the dynamics of velocity selection on diffraction phase systematics in a Bragg-diffraction-based Ramsey-Bordé atom interferometer. By tuning the intensity-dependent differential light shift during velocity-selective Raman transitions, we demonstrate experimentally that the intensity of the velocity selection laser—a state preparation step rather than part of the interferometry sequence—significantly alters the final diffraction phase landscape. Therefore, achieving milliradian-level control of the diffraction phase requires a model of the velocity selection process integrated into the numerical interferometer simulation. We have identified a “magic detuning” for velocity selection at which the resonant velocity condition becomes insensitive to the Raman laser intensity.  Operating at this detuning makes the diffraction phase landscape effectively insensitive to Raman laser intensity and eliminates spatial velocity variation accross the atom cloud. Although the precise value of this detuning is sensitive to experimental parameters, we demonstrate that the magic detuning can be readily identified using several independent experimental approaches. This approach simplifies the numerical modeling required to correct diffraction phase systematics and offers a technique for constructing atom interferometers that are robust to drifts in the velocity selection laser system.
\vspace{0.5cm}
\section{Acknowledgement}
\vspace{-0.25cm}
This research was supported by U.S. National Science Foundation grants 2208029, 2512532, 1806583, and 2328663 through Subaward No. 3068 from Rutgers University; National Aeronautics and Space Administration (NASA) grant 80NSSC25K7641 via subaward SUB00001274 through the University of Rochester; Heising-Simons Foundation grant 2026-6565; Orolia Defense and Security, LLC (dba Safran Federal Systems Inc.) under Agreement No. 062257; and Gordon and Betty Moore Foundation grant 9366. Y.I. is supported by the NSF Graduate Research Fellowship and the Paul and Daisy Soros Fellowship for New Americans. 



\bibliography{mwl_paper_bib}

\end{document}